\documentclass[aps,onecolumn,pra,showpacs,notitlepage,superscriptaddress,floatfix]{revtex4-1}
\usepackage[dvipsnames]{xcolor}

\usepackage{amssymb,amsmath,amsfonts}
\usepackage{epsfig,graphicx}
\usepackage{hyperref}

\hypersetup{colorlinks,citecolor=PineGreen,filecolor=black,linkcolor=black,urlcolor=black}
\usepackage{epstopdf}
\usepackage{subfigure}

\usepackage{color}

\newcommand{\adag}{a^{\dagger}}
\newcommand{\adaga}{a^{\dagger}a}
\newcommand\ket[1]{\left|\textstyle{#1}\right\rangle}

\begin{document}

\title{Magnetic field fluctuations analysis  for the ion trap implementation  of the quantum Rabi model in the deep strong coupling regime}

\author{Ricardo Puebla}\affiliation{Institut f\"{u}r Theoretische Physik and IQST, Albert-Einstein-Allee 11, Universit\"{a}t Ulm, D-89069 Ulm, Germany}
\author{Jorge Casanova}\affiliation{Institut f\"{u}r Theoretische Physik and IQST, Albert-Einstein-Allee 11, Universit\"{a}t Ulm, D-89069 Ulm, Germany}
\author{Martin B. Plenio}\affiliation{Institut f\"{u}r Theoretische Physik and IQST, Albert-Einstein-Allee 11, Universit\"{a}t Ulm, D-89069 Ulm, Germany}

\begin{abstract}
The dynamics of the quantum Rabi model in the deep strong coupling regime  is theoretically analyzed in a trapped-ion setup. Recognizably, the main hallmark of this regime is the emergence of collapses and revivals, whose faithful observation is hindered under realistic magnetic dephasing noise. 
Here we discuss how to attain a faithful implementation of the quantum Rabi model in the deep strong coupling regime which is robust against magnetic field fluctuations and at the same time provides a large tunability of the simulated parameters. This is achieved by combining standing wave laser configuration with continuous dynamical decoupling. In addition, we study the role that amplitude fluctuations play to correctly attain the quantum Rabi model using the proposed method. In this manner the present work further supports the suitability of continuous dynamical decoupling techniques in trapped-ion settings to faithfully realize different interacting dynamics.
\end{abstract}


\maketitle

\section{Introduction}
\label{sec:int}
One of the simplest, yet fundamental, quantum models consists of a two-level system interacting with a single-mode bosonic field.  This system, besides of being a nice textbook example of fundamental quantum physics, describes realistic phenomena in a variety of physical situations, and it is commonly known as quantum Rabi model (QRM), in honor of the groundbreaking work of the author of the same name who analyzed the interaction of a spin with a classical field~\cite{Rabi:36,Rabi:37}.   Certainly, this simple model emerges naturally in different physical situations; although primarily studied in the realm of quantum optics, its relevance encompasses even quantum information processing~\cite{Romero:12}. This underlies the considerable attention that this model has attracted and the efforts devoted during the last decades to elucidate the  physics of the QRM~\cite{Braak:16}. Furthermore, although the first fully quantized version of this model dates from 1963~\cite{Jaynes:1963fa}, the QRM  still reveals new results, such as its integrability~\cite{Braak:2011hc}, the appearance of a finite-component quantum phase transition in an appropiate limit~\cite{Hwang:2015eq,Puebla:2016esqpt} or the structured dynamics of revivals in the deep strong coupling (DSC) regime~\cite{Casanova:2010kd} to name a few recent insights. 

However, the experimentally accessible parameter regime of a light-matter interacting system, where the QRM naturally emerges, is generally constrained to small couplings between the two subsystems (qubit and bosonic mode), hindering the observation of the rich phenomenology of the ultrastrong coupling regime. In this respect, recent works paved the way to scrutinize the ultrastrong coupling regime of the QRM, as in cavity QED~\cite{Niemczyk:10,Forn:10} or in a spin-mechanical system~\cite{Abdi:17}. Nevertheless, with the advent of quantum simulation techniques, exploring aspects of strong light-matter interaction is now possible in diverse platforms, which are unfeasible or hardly attainable otherwise. Among the diverse setups where the QRM can be realized, trapped ions merit special attention as they combine qubits possessing long coherence time with high fidelity measurements~\cite{Leibfried:03r,Leibfried:03}. Furthermore, an implementation of the QRM in a large variety of parameter regimes is possible in a trapped-ion setup~\cite{Pedernales:15}, which enables the exploration of the DSC regime or the emergence of the quantum phase transition~\cite{Puebla:17}. However, an accurate realization of the QRM crucially depends on the mitigation of the impact of different experimental imperfections into trapped-ion dynamics. We study this situation in a trapped-ion setup in which magnetic field fluctuations constitute the main source of decoherence, that is, these fluctuations limit the coherence time of the system. Yet, although another noise sources may be also relevant, such as laser phase noise or motional decoherence, we consider here that their impact produces decoherence in a larger time scale than magnetic field fluctuations,  as it is the case of the experiments performed involving metastable states of optical ions~\cite{Gerritsma:10,Gerritsma:11} or microwave-driven ions~\cite{Timoney:11,Piltz:16}. In addition to magnetic-dephasing noise, we will discuss also the effect of fluctuations in the amplitude of the lasers. Hence, the realization of a QRM with a trapped ion may be enhanced by means of schemes that are robust against these noise sources.  In this context, the suitability of dynamical decoupling (DD) techniques to simulate a QRM in a trapped ion setting has been recently reported by the authors in~\cite{Puebla:16}, and constitutes a promising tool to attain prolonged coherence times in setups mainly affected by magnetic-dephasing noise. However, the particular parameter regime of the QRM may not be trivially achieved with DD techniques due to either a breakdown of required approximations or the impossibility  to have access to the  desired parameters. Certainly, this fact challenges the achievement of a protected QRM in the DSC regime. 

In this work we show that by combining continuous dynamical decoupling methods and standing wave configuration of the lasers to create interaction terms~\cite{Cirac:92,deLaubenfels:15}, a magnetic-dephasing noise-resilient QRM can be accomplished even in the DSC regime. We illustrate the advantage of this scheme with respect to a bare realization by means of numerical simulations to observe the main hallmark of the QRM in this regime, namely, the emergence of collapses and revivals in the dynamics~\cite{Casanova:2010kd}. The article is organized as follows. We first introduce the QRM in Sec.~\ref{sec:QRM}, and comment the physics of the DSC regime. In Sec.~\ref{sec:DD} we first introduce the trapped-ion setup and discuss the standard method to realize a QRM without protection against magnetic field fluctuations. Then, we show that a magnetic-dephasing noise resilient, yet tunable, QRM in can be obtained in a trapped-ion setup combining both continuous dynamical decoupling methods and standing wave configuration, even in the presence of amplitude fluctuations, as demonstrated by numerical simulations. Finally, in Sec.~\ref{sec:sum} we summarize the main results.
 
\section{Quantum Rabi model}
\label{sec:QRM}
The QRM describes the interaction of a two-level system with a single-mode bosonic field. The Hamiltonian of this  model can be written as
\begin{equation}
\label{eq:QRM}
H_{\rm QRM}=\frac{\omega}{2}\sigma_z+\omega_0 \adaga-\lambda \sigma_x\left( a+\adag\right),
\end{equation}
where $\sigma_{x,y,z}$ are the Pauli operators acting on the two-level system, whose transition frequency is $\omega$, and the usual creation and annihilation bosonic operators ($a^\dag$ and $a$) give account of the quantized single-mode bosonic field with frequency $\omega_0$. The subsystems interact through the last term of the Hamiltonian, with strength given by the coupling constant $\lambda$. It is worth mentioning that the Hamiltonian includes the so-called counter-rotating terms, namely, $\sigma^+\adag$ and $\sigma^-a$, which do not appear in the Jaynes-Cummings model (JC)~\cite{Jaynes:1963fa}. As a consequence, the QRM does not conserve the total number excitations in the system, $N=\adaga+\sigma^+\sigma^-$. Nonetheless, the parity of $N$ is conserved, namely this $Z_2$ parity symmetry has an associated  operator, $\Pi=-\sigma_z(-1)^{\adaga}$, that commutes with the Hamiltonian and allows us to gain insight in its system dynamics~\cite{Casanova:2010kd}. 

In this work we focus in the DSC regime of the QRM. This regime takes place beyond the strong or ultrastrong coupling regimes~\cite{Niemczyk:10, Forn:10}, that is, when the coupling constant becomes equal or larger than the bosonic frequency $\lambda/\omega_0\gtrsim 1$~\cite{Casanova:2010kd, Rossatto:16}. The dynamics in this regime is characterized by collapses and revivals, which are periodically occurring in the solvable case $\omega=0$~\cite{Casanova:2010kd}. As demonstrated analytically in the latter case, an initial state $\ket{\psi(0)}$ evolves drawing periodic orbits in the phase space. Hence, the survival probability $S(t)=\left|\left< \psi(0)|\psi(t) \right> \right|^2$ reaches $1$ at times $t_m= m 2\pi/\omega_0$ with $m$ a positive integer number, meaning that the initial state is perfectly retrieved. However, this situation slightly differs when considering a more general case, $\omega\neq 0$ but  $\omega\sim \omega_0$, where despite the dynamics is reminiscent of that of $\omega=0$, collapses and revivals are hindered by a non-regular distribution of the energy spectrum. Finally, we remark that the off-resonant condition $\omega\gg\omega_0$ becomes particularly interesting due to the emergence of a quantum phase transition~\cite{Hwang:2015eq}.

\begin{figure}
\centering
\includegraphics[width=0.7\linewidth,angle=-00]{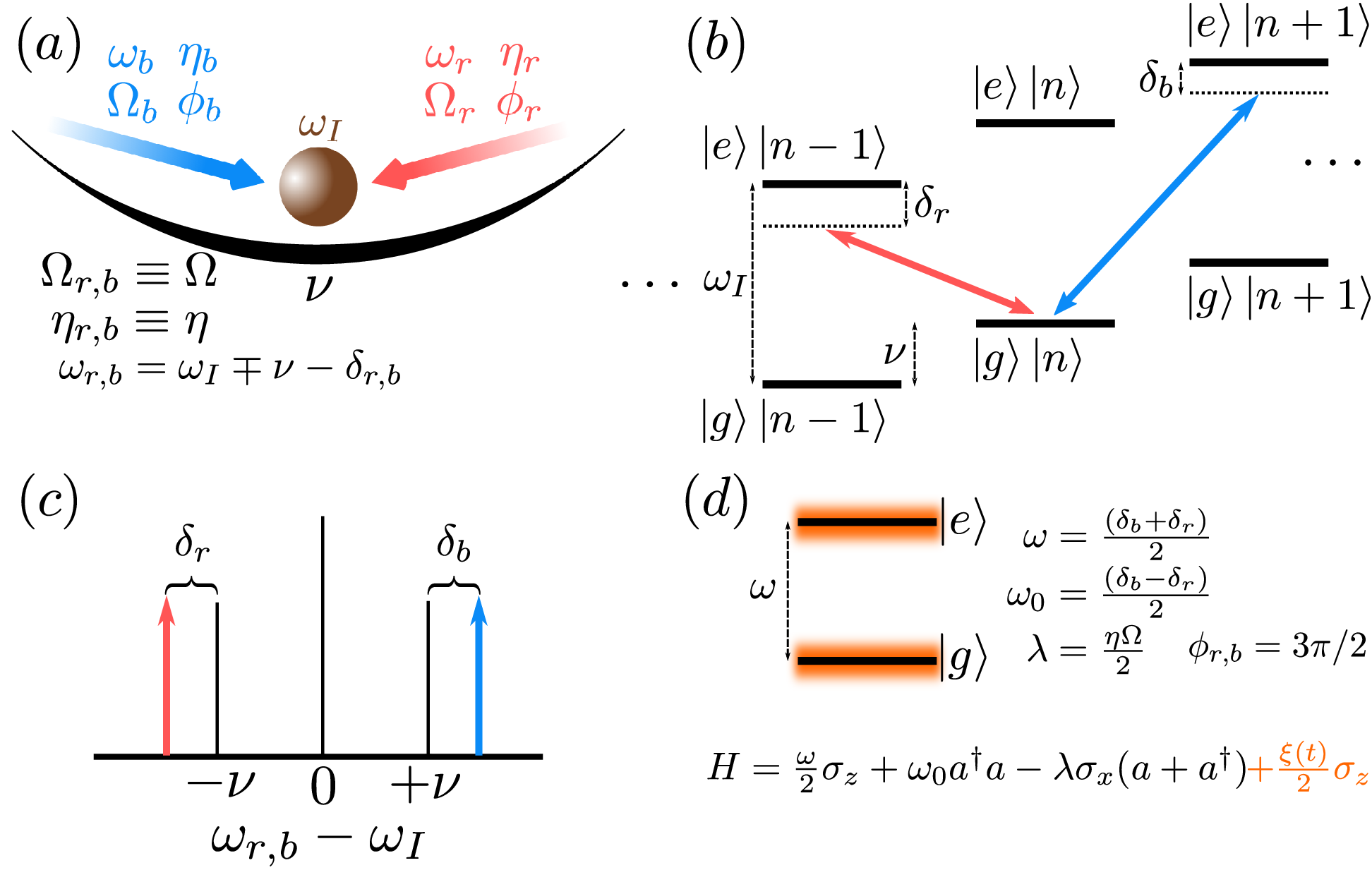}
\caption{ Illustration of the standard scheme to attain a QRM with a trapped-ion setup. In (a) we sketch the ion trapped in a harmonic potential of frequency $\nu$ and irradiated by two lasers, driving detuned red- and blue-sidebands, characterized by frequency, amplitude, Lamb-Dicke and phase, $\omega_{r,b}$, $\Omega_{r,b}$, $\eta_{r,b}$ and $\phi_{r,b}$, respectively. They generate the processes  depicted in (b) by red and blue arrows, where $\ket{e}$ and $\ket{g}$ refer to the ion internal states and $\ket{n}$ represents the motional Fock states. In (c) we show the corresponding laser tones in frequency space, with respect to the qubit splitting $\omega_I$. Finally, in (d) the parameters of the simulated QRM are given. However, the magnetic-dephasing noise spoils a correct realization of the QRM.} \label{fig:scheme}
\end{figure}

\section{Trapped-ion realization of the QRM in the DSC regime}
\label{sec:DD}
In the following we analyze the impact of magnetic field fluctuations on the realization of a QRM in the DSC regime using a trapped ion and how these fluctuations can be reduced by means of continuous DD techniques. For that, we firstly review the standard scheme to accomplish a QRM and discuss how to incorporate magnetic field fluctuations into system dynamics. Secondly, we propose a continuous DD scheme which together with a standing wave configuration allow us to realize the QRM in this regime, as demonstrated by the numerical simulations.

We consider a typical trapped-ion scenario where the qubit is encoded using metastable states with optical transition denoted here as $\omega_I$. In experiments performed with ${}^{40}{\rm Ca}^{+}$ ions the transition is $\omega_I\approx2\pi \times 4\cdot 10^{14}$ Hz~\cite{Gerritsma:10,Gerritsma:11}. The dynamics of the ion, confined in a trap with center-of-mass frequency $\nu$, interacting with a set of lasers is dictated by the following Hamiltonian~\cite{Leibfried:03r}
\begin{equation}
\label{eq:HTI}
H_{\rm TI}=\frac{\omega_I}{2}\sigma_z+\nu \adaga +\sum_j \Omega_j \sigma_x \cos\left( \vec{k}_j\cdot \vec{x}-\omega_jt-\phi_j\right),
\end{equation}
where amplitude, wave vector, frequency and initial phase of each laser is denoted by $\Omega_j$, $\vec{k}_j$, $\omega_j$ and $\phi_j$, respectively. Considering motion along one direction, $\vec{x}=x$, the position can be written in terms of the corresponding vibrational mode, $x=x_0 (a+\adag)$ with $x_0=1/\sqrt{2m\nu}$ and $m$ the ion mass. The coupling between the motional states and internal electronic states of the ion is quantified by the Lamb-Dicke parameter, $\eta_j=k_j x_0$. In this setup, an optical driving with $\omega_j\approx \omega_I$ can provide with a reasonably large Lamb-Dicke parameter  $\eta\approx 0.04$~\cite{Gerritsma:11} and Rabi frequency $\Omega_j\sim 10^1-10^{2}$ kHz, while the trap frequency $\nu$ takes values on the range of $1$ MHz. 

We consider that the internal states of the ion, $\ket{g}$ and $\ket{e}$,  are magnetically sensitive, as it is the case of a setup with ${}^{40}{\rm Ca}^{+}$ ions~\cite{Gerritsma:10,Gerritsma:11}. In these setups, coherence time is typically limited by magnetic field fluctuations, although different noise sources, such as amplitude fluctuations, phase noise or heating of the motional mode, may also be relevant. However, while these noise sources certainly entail decoherence, we consider that their significant impact will occur at longer evolution times. Indeed, heating rate and laser phase noise can be estimated as $1$ phonon/s~\cite{Brownnutt:15} and $1$ rad/s~\cite{Walther:12}, respectively, while the typical  coherence time of these systems lies in the range of milliseconds. Thence, we constrain ourselves to the analysis of magnetic field fluctuations, although the impact of amplitude fluctuations and further imperfections will be discussed later on, and incorporated when presenting the DD scheme. Moreover, since the lifetime of the qubit states is much longer than the time it takes to lose phase coherence between them, i.e., $T_1\gg T_2$, and because the evolution times considered here are $\sim T_2$, only longitudinal magnetic field fluctuations are included. Note that, while $T_2$ takes values in the order of few milliseconds, $T_1$ can be larger than seconds.  In particular,  for our numerical simulations we take $T_2= 3$ ms according to~\cite{Gerritsma:11}. It is worth emphasizing that this is also the case for setups involving microwave-driven ions~\cite{Timoney:11,Piltz:16}, where $T_2\approx 5$ ms is observed by using the hyperfine levels of the  ${}^{171}{\rm Yb}^{+}$ ion~\cite{Timoney:11}. The effect of such magnetic field fluctuations can be effectively captured by adding a term $\xi(t)\sigma_z/2$ to the trapped-ion Hamiltonian, such that $\xi(t)$ represents a stochastic process~\cite{Bermudez:12,Cai:12,Lemmer:13,Mikelsons:15,Puebla:16,Puebla:17dd}. Such a fluctuation is commonly described as an Orstein-Uhlenbeck process with zero mean~\cite{Uhlenbeck:30,Gillespie:96,Gillespie:96A}, which is a Markovian and Gaussian process exhibiting a finite-width spectral density. Thus, the time evolution of $\xi(t)$ depends on two parameters, namely, relaxation time $\tau$ and diffusion constant $c$, which determine the total noise power $P_{\rm MF}=c\tau/2$. Note that $P_{\rm MF}\equiv C_{\xi}(0)$ as it follows from the Wiener-Khinchin theorem, where $C_\xi(t')$ is the auto-correlation function, $C_{\xi}(t')=\left< \xi(t)\xi(t+t')\right>$~\cite{Gillespie:96,Gillespie:96A}. The relaxation time determines the width of the spectral density, $1/(2\pi\tau)$, and it can be estimated to be $\tau\approx100\ \mu$s in trapped-ion experiments~\cite{Mikelsons:15}. The diffusion constant is then determined to correctly reproduce the coherence time $T_2$, which for the typical trapped-ion scenario, $T_2\gg \tau$, simplifies to $c\approx 2/(\tau^2 T_2)$~\cite{Bermudez:12,Lemmer:13}, and thus, it exhibits a noise power $P_{\rm MF}=1/(\tau T_2)$. We remark that this noise model successfully reproduces the coherence decay observed experimentally~\cite{Wineland:98,Cai:12}.

\subsection{Standard scheme}

The trapped-ion Hamiltonian, Eq.~(\ref{eq:HTI}), allows to simulate the QRM in an appropriate interaction picture by driving detuned red- and blue-sidebands (denoted with the subscripts $r$ and $b$, respectively), which provides with the needed JC and anti-JC interaction terms of the QRM~\cite{Cirac:93,Pedernales:15}. The frequencies of these two lasers are set to $\omega_{r,b}=\omega_I\mp \nu -\delta_{r,b}$, with $\delta_{r,b}\ll \nu\ll \omega_I$. See Fig.~\ref{fig:scheme} for an illustration of this scheme.  Hence, $|\omega_I-\omega_{r,b}|\ll |\omega_I+\omega_{r,b}|$ which together with the condition $\eta\sqrt{\left<(a+\adag)^2\right>}\ll 1$ (Lamb-Dicke regime), the trapped-ion Hamiltonian in an interaction picture with respect to $H=\omega_I/2\sigma_z +\nu\adaga$, adopts the following form~\cite{Cirac:93,Pedernales:15}, 
\begin{equation}
H^I_{\rm TI}\approx \frac{\xi(t)}{2}\sigma_z-\frac{\eta\Omega}{2}\left[\sigma^+a e^{-i\delta_rt} +\sigma^+\adag e^{-i\delta_bt}+\rm{H.c.} \right],
\end{equation}
with $\Omega\equiv\Omega_{r,b}$, $\eta\equiv \eta_{r,b}$, $\phi_{r,b}=3\pi/2$, and where the magnetic-fluctuation term, $\xi(t) \sigma_z/2$, is explicitly shown. Indeed, the previous Hamiltonian corresponds to a QRM, Eq.~(\ref{eq:QRM}), in a rotating frame with respect to the free terms $\omega/2\sigma_z+\omega_0\adaga$, where simulated parameters of the QRM are $\omega=(\delta_b+\delta_r)/2$, $\omega_0=(\delta_b-\delta_r)/2$ and the coupling constant $\lambda=\eta\Omega/2$. However, the magnetic-dephasing noise spoils a correct realization of a QRM. It is therefore pertinent to develop a scheme to cope with these magnetic field fluctuations.

\subsection{Continuous dynamical decoupling scheme}

\begin{figure}
\centering
\includegraphics[width=0.7\linewidth,angle=-00]{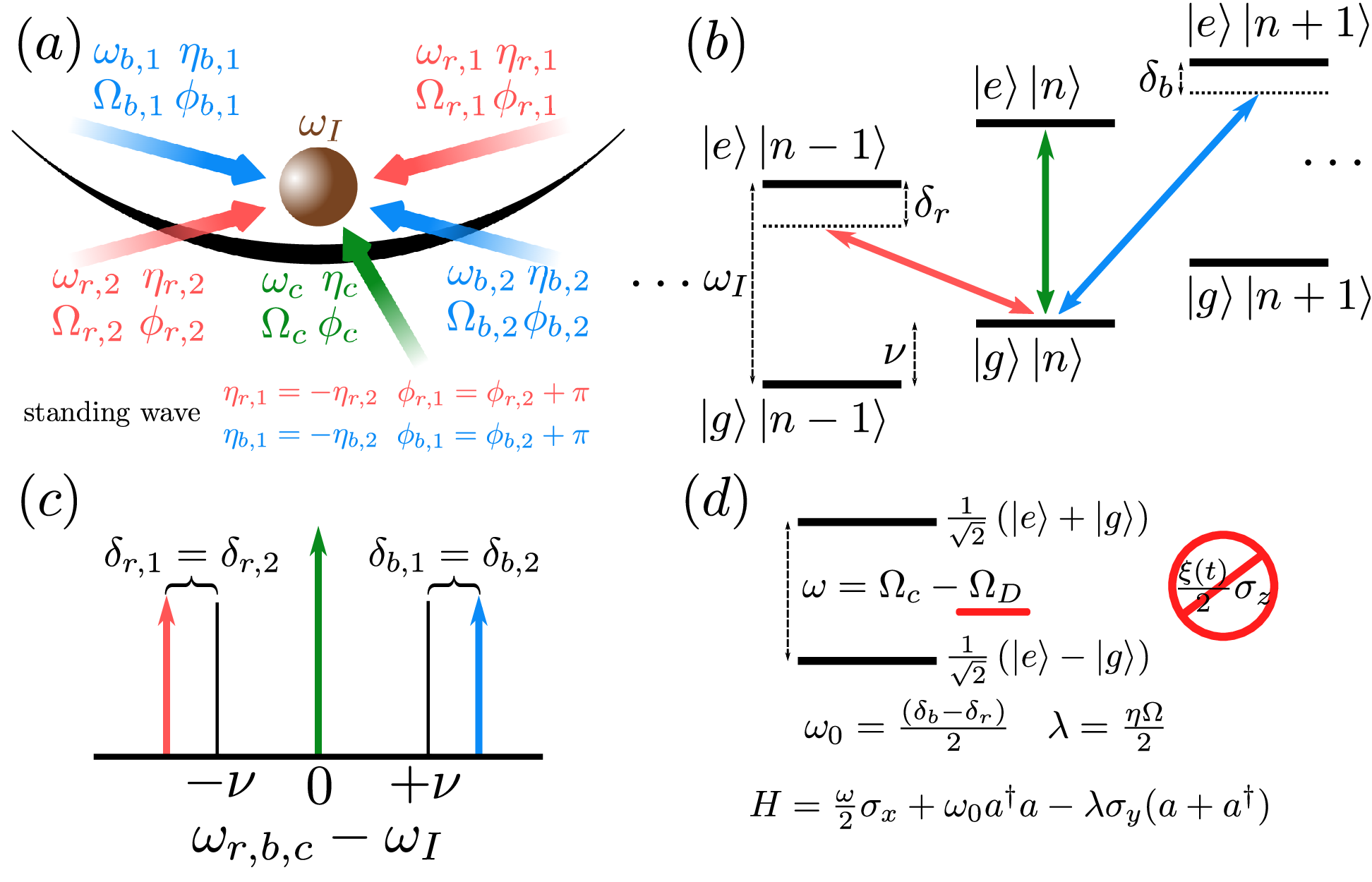}
\caption{ Illustration of the continuous DD scheme for the implementation of a QRM in a trapped ion setup. In (a) we sketch the ion trapped in a harmonic potential of frequency $\nu$ and irradiated by the required lasers, which drive two detuned red- and blue-sidebands in a standing-wave configuration and a carrier, represented by red, blue and green arrows, respectively. In (b) we show the processes driven by these lasers, while in (c) we show their corresponding laser tones in frequency space, with respect to the qubit splitting $\omega_I$. Note that the two red (blue) detuned sidebands exhibit the same detuning. Finally, in (d) we show the dressed basis in which the QRM is realized, as well as the parameters of the QRM. As a consequence of the dressed basis provided by the carrier interaction, the original magnetic field fluctuations affecting the trapped ion  are averaged out, and thus a protected QRM can be realized (see main text for further details and Table~\ref{tab:1} regarding how the trapped-ion setup parameters, $\omega_j$, $\delta_j$, $\Omega_j$, $\eta_j$ and $\phi_j$, relate to the simulated QRM parameters.)} \label{fig:schemeSW}
\end{figure}

A continuous DD scheme can be used to realize the QRM while it overcomes the aforementioned magnetic-dephasing noise, as recently proposed in~\cite{Puebla:16}. However, the realization of a QRM in the DSC regime with $\omega=0$ or $\omega\leq\omega_0$ can not be trivially achieved within the reported scheme, and demands a different strategy, as we discuss below. In general, a continuous DD method consists in providing with a new qubit dressed basis by means of an additional driving.  Then, as the noise term becomes orthogonal to the new qubit basis, its effect is mostly suppressed  if the corresponding qubit frequency splitting is sufficiently large such that the noise is not capable to produce transitions.
In our case, the dressed basis is accomplished by an additional laser, driving a carrier interaction denoted by a subscript $c$, in addition to the detuned red- and blue-sidebands. Then, the resulting Hamiltonian in an interaction picture with respect to $H=\omega_I/2\sigma_z +\nu\adaga$ reads
\begin{equation}
H^{I}_{\rm TI}\approx \frac{\xi(t)}{2}\sigma_z+\frac{\Omega_c}{2}\sigma_x+\frac{\eta\Omega}{2}\left[ \sigma^+a e^{i\delta_rt}e^{-i(\phi_r-\pi/2)}+\sigma^+\adag e^{i\delta_bt}e^{-i(\phi_b-\pi/2)}+{\rm H.c}\right].
\end{equation}
Note that the previous Hamiltonian is achieved, first neglecting terms rotating at frequency $\omega_{c,r,b}+\omega_I$ (optical RWA) and those at $\nu$ or higher, within the Lamb-Dicke regime. These approximations  can be safely performed as in the standard case.  In addition, for the carrier interaction, $\omega_c=\omega_I$, the initial phase has been already set to $\phi_c=0$, while its corresponding Lamb-Dicke parameter $\eta_c$ is not relevant since all of its sidebands are detuned by an amount $\approx |\nu|$ and thus, are averaged out.

\begin{table}\label{tab:1}
  \begin{center}
    \begin{tabular}{ccccc} \hline
      
      Carrier & \multicolumn{2}{c}{Detuned red sidebands} & \multicolumn{2}{c}{Detuned blue sidebands} \\ \cline{1-5} 
  $\omega_c=\omega_I$ & $\omega_{r,1}=\omega_I-\nu-\delta_{r,1}$ & $\omega_{r,2}=\omega_I-\nu-\delta_{r,2}$  & $\omega_{b,1}=\omega_I+\nu-\delta_{b,1}$  &$\omega_{b,2}=\omega_I+\nu-\delta_{b,2}$  \\ 
  $\delta_c=0$ & $\delta_{r,1}=\Omega_D-\omega_0$ &$\delta_{r,2}=\Omega_D-\omega_0$ &$\delta_{b,1}=\Omega_D+\omega_0$ &$\delta_{b,2}=\Omega_D+\omega_0$ \\
 $\Omega_c=\Omega_D+\omega$ & $\Omega_{r,1}=\Omega$ & $\Omega_{r,2}=\Omega$ & $\Omega_{b,1}=\Omega$ & $\Omega_{b,2}=\Omega$ \\
     $\eta_c$ & $\eta_{r,1}=\eta$ & $\eta_{r,2}=-\eta$ & $\eta_{b,1}=\eta$ & $\eta_{b,2}=-\eta$  \\
     $\phi_c=0$ & $\phi_{r,1}=0$ & $\phi_{r,2}=\pi$ & $\phi_{b,1}=0$ & $\phi_{b,2}=\pi$ \\\hline
\end{tabular}\caption{\small{Trapped-ion parameters for the continuous DD scheme using a standing-wave configuration, as dicussed in the main text.  The parameters of the simulated QRM are $\omega=\Omega_c-\Omega_D$, $\omega_0=(\delta_{b,1}-\delta_{b,1})/2$ and $\lambda=\eta\Omega/2$ with a rotated spin basis with respect to Eq.~(\ref{eq:QRM}).}}
\end{center}
\end{table}

As a consequence of the carrier interaction, a new dressed basis can be defined according to $\Omega_c/2 \sigma_x$. Then, it can be shown that the noisy term $\xi(t) \sigma_z/2$ does not produce spurious excitations in the new dressed basis if $\Omega_c> 1/(2\pi\tau)$ and provided that it is not too strong. In a trapped-ion case,  with $T_2=3$ ms and $\tau=100\ \mu$s, a Rabi frequency  of the carrier $\Omega_c\sim 10^1-10^2$ kHz can be sufficient to overcome these fluctuations, and hence, the term $\xi(t)$ can be safely neglected~\cite{Cai:12,Puebla:16,Puebla:17dd}. Therefore, the QRM could be readily achieved from the previous Hamiltonian with parameters $\omega=\Omega_c$, $\omega_0=-\delta_r=\delta_b$ and $\lambda=\eta\Omega/2$~\cite{Puebla:16}, although the accessible parameter regime would be unnecessarily constrained to a large value of the qubit frequency in the QRM, $\omega$ (for an efficient elimination of the noise). Therefore, in order to gain tunability of the simulated parameters,  we propose here a different method, which is sketched in Fig.~\ref{fig:schemeSW} and explained in the following. We first move to an additional rotating frame with respect to $\Omega_{D} \ \sigma_x/2$ such that $\Omega_c=\Omega_{D}+\omega$, that results in the following  Hamiltonian
\begin{align}
H^{II}_{\rm TI}\approx \frac{\omega}{2}\sigma_x&+\frac{\eta \Omega}{4} \left[ \left(i\sigma_x-\cos(\Omega_Dt)\sigma_y+\sin(\Omega_Dt)\sigma_z \right)\times\right.\nonumber\\&\times\left.\left(a e^{i\delta_rt}e^{-i\phi_r} +\adag e^{i\delta_b t}e^{-i\phi_b}\right)+{\rm H.c.}\right],
\end{align}
where the noisy term has been already neglected as expected for large enough values of $\Omega_{D}$ (see later for a numerical confirmation). Finally, choosing detunings such that $\delta_{r,b}=\Omega_D\mp \omega_0$ and initial phases $\phi_{r,b}=0$, and performing an extra rotating-wave approximation to eliminate terms rotating at $\Omega_D$ or higher frequencies, the Hamiltonian adopts the form of a QRM in the rotating frame of $\omega_0\adaga$, i.e.
\begin{equation}
H^{II}_{\rm TI}\approx \frac{\omega}{2}\sigma_x-\frac{\eta \Omega}{4} \sigma_y\left[a e^{-i\omega_0t} +\adag e^{i\omega_0t}\right],
\end{equation}
where the spin basis is now rotated with respect to that of the QRM given in the Eq.~(\ref{eq:QRM}). Note however that the previous RWA is only valid if $\Omega_D\gg\eta\Omega$, but with $\Omega_D$ still small compared to the trap frequency $\nu$ to properly resolve sidebands, $|\Omega_D\mp\omega_0|\ll\nu$, and at the same time $\Omega_{r,b}\ll |\Omega_D-\nu|$ to safely neglect off-resonant carrier interactions generated by the red- and blue-sidebands. Indeed, although the impact of the magnetic-dephasing noise is largely reduced, the QRM is not correctly achieved due to the failure of the latter RWA (i.e. the one aimed to eliminate the detuned carrier contribution). For typical parameters  $\nu\sim $ MHz and $\Omega_{r,b,D}\sim 10^1-10^2$ kHz, these detuned carrier interactions become relevant. 
\begin{figure}[t]
\centering
  \includegraphics[width=0.475\linewidth,angle=-90]{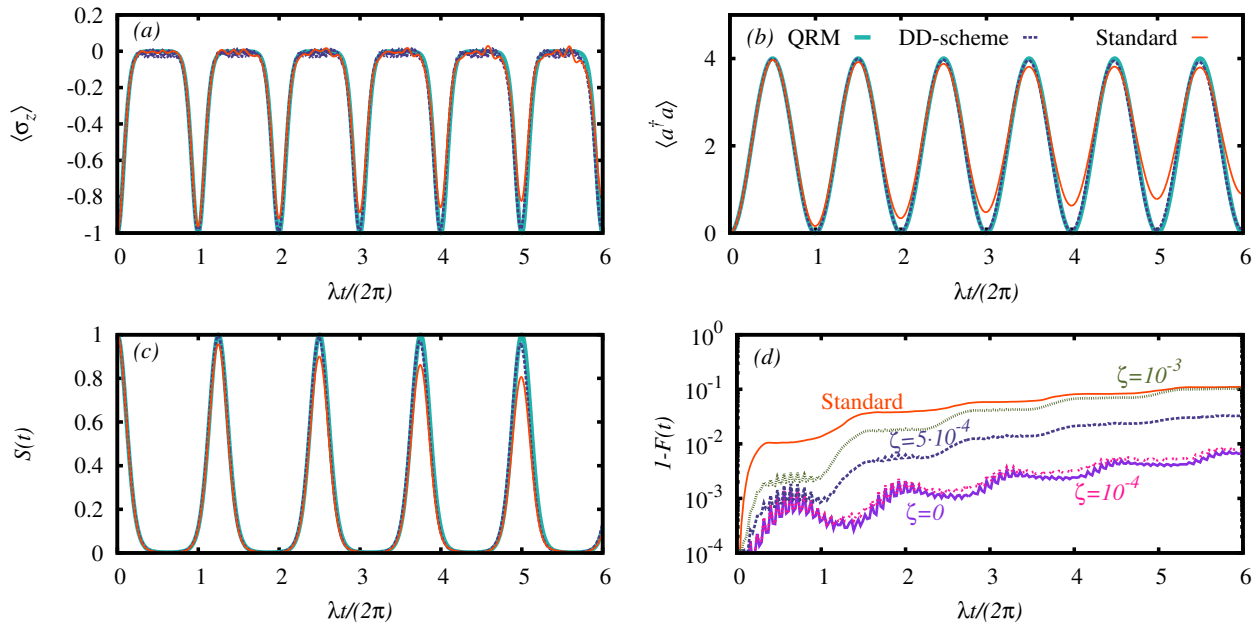}
\caption{ Trapped-ion dynamics of the QRM with $\omega=0$ in the DSC regime where collapses and revivals emerge. The ideal dynamics is plotted with a solid light-blue line, the unprotected (standard) realization with solid red lines and the protected scheme with dotted dark blue lines, which include amplitude fluctuations with $\zeta=5\times 10^{-4}$. In (a) and (b) the simulated QRM coupling constant is $\lambda/\omega_0=1$ and the initial state $\ket{\psi(0)}=\ket{g}\ket{0}$, according to Eq.~(\ref{eq:QRM}), where we show the time evolution of spin and phonon population, namely, $\left<\sigma_z \right>$ and $\left<\adaga \right>$.   In (c) and (d) the initial state  $\ket{\psi(0)}=1/\sqrt{2}(\ket{e}-\ket{g})\ket{2}$ evolves under a simulated QRM with $\lambda=1.25\omega_0$. The survival probability, $S(t)=\left|\left< \psi(0)|\psi(t) \right> \right|^2$, is plotted in (c),  while (d) corresponds to the infidelity, $1-F(t)=1-\left|\left<\psi_{\rm QRM}(t) \right|\left.\psi(t) \right>\right|$ where $\ket{\psi_{\rm QRM}(t)}$ stands for the quantum state evolved under the ideal QRM and $\ket{\psi(t)}$ for its trapped-ion counterpart, by means of the standard  scheme (red) and for the DD scheme with different $\zeta$ values, i.e., with different strength of amplitude fluctuations. The trapped-ion results were obtained after an ensemble average of $100$ stochastic trajectories, and assuming only the optical RWA. The impact of magnetic-dephasing noise into the bare realization is significant, which is accentuated at the end of the evolution, after $6$ ms since $\lambda=2\pi\times 1$ kHz. As expected, the method based on the DD scheme exhibits a similar performance when amplitude fluctuations become strong, namely, when $P_{\rm MF}\approx P_{\rm AF}$, which for the parameters considered here occurs for $\zeta\approx 10^{-3}$, while for weaker amplitude fluctuations the DD scheme becomes favorable since magnetic-dephasing noise is overcome. See main text for further details as well as information regarding simulation parameters.} \label{fig:1}
\end{figure}

Therefore, on top of the above commented DD scheme, we propose a standing wave configuration of the red- and blue-sidebands to eliminate the off-resonant carrier interactions~\cite{Cirac:92,deLaubenfels:15,Puebla:17}, while at the same time driving the carrier with a traveling wave. This consists in placing the ion at the node of two counter-propagating waves, that is, $\phi_{r,1}=\phi_{r,2}+\pi$ and $\vec{k}_{r,1}\cdot\vec{x}=-\vec{k}_{r,2}\cdot\vec{x}$ or equivalently, $\eta\equiv\eta_{r,1}=-\eta_{r,2}$. In this manner, the spurious off-resonant carrier interaction generated by the detuned sidebands is canceled. Note that two lasers are now employed to drive the detuned red-sideband, $\omega_{r,1}=\omega_{r,2}=\omega_I-\nu-\Omega_D+\omega_0$, and similarly for the blue-sideband, which finally lead to
\begin{equation}
  \label{eq:HSW}
H^{II,SW}_{\rm TI}\approx \frac{\omega}{2}\sigma_x-\frac{\eta \Omega}{2} \sigma_y\left[a e^{-i\omega_0t} +\adag e^{i\omega_0t}\right],
\end{equation}
where $\Omega=\Omega_{r,j}=\Omega_{b,j}$ and $\phi_{r,1}=\phi_{b,1}=0$. The previous Hamiltonian corresponds to a QRM rotating in the frame of $\omega_0\adaga$, and with parameters $\omega=\Omega_c-\Omega_D$, $\omega_0=(\delta_{b}-\delta_r)/2$ and $\lambda=\eta\Omega/2$. This method is sketched in Fig.~\ref{fig:schemeSW}, where the required lasers are described  and the corresponding dressed basis in which the qubit is encoded. In addition, the trapped-ion parameters are collected in Table~\ref{tab:1}, as well as their relation to those of the simulated QRM.
In this manner we can effectively eliminate the magnetic-dephasing noise without spurious carrier contributions and, at the same time, define the parameters of the simulated QRM (note that $\omega=\Omega_c-\Omega_D$). This strategy allows us to explore a wide parameter regime including the DSC, as $\omega$ can be set now to the regime $\omega \lesssim \omega_0 \lesssim \lambda$, improving the standard realization even in the presence of further realistic imperfections, as we comment in the following and support by means of numerical simulations.

While magnetic-dephasing noise can be overcome by means of the proposed scheme, the performance of this method can be limited by other realistic imperfections in its implementation. Indeed, within this scheme, amplitude fluctuations in the carrier tone may become now the main source of noise since it provides the dressed basis used to simulate the QRM.   As aforementioned, other noise sources are expected to have an impact at longer evolution times. In addition, it is also worth mentioning that since both schemes rely on the same motional degree of freedom and the QRM is accomplished in the same time, motional decoherence processes will deteriorate similarly both schemes.
  Amplitude fluctuations can be modeled as $\Omega_c\rightarrow \Omega_c(1+\beta(t))$ where $\beta(t)$ represents a noise with zero mean and relative strength $\zeta$, that is, $\zeta=0.01$ indicates a fluctuation of a $1\%$ of the amplitude $\Omega_c$, and thus, it features a total noise power $P_{\rm AF}=(\zeta\Omega_c)^2$.  As for magnetic field fluctuations, we model $\beta(t)$ as an Orstein-Uhlenbeck process but with a longer relaxation time $\tau_\beta=1$ ms~\cite{Lemmer:13,Cai:12,Puebla:16}. This realistic noise may challenge the performance of the proposed method with respect to the unprotected standard scheme. Indeed, if amplitude fluctuations surpass the effect of the original magnetic-dephasing noise, the proposed method becomes counter-productive. This disadvantageous scenario is expected to occur when $P_{\rm AF}\geq P_{\rm MF}$, which within the explained noise model, it leads to $\zeta\Omega_c\geq 1/\sqrt{\tau T_2}$.  We will show by means of numerical simulations the impact of this additional noise source including a fluctuating term $\Omega_c(1+\beta(t))$ with different $\zeta$ values. We remark that although all the lasers may suffer amplitude fluctuations, the lasers driving red- and blue-sidebands produce a much smaller impact in the Hamiltonian than the carrier interaction since they are accompanied by the corresponding Lamb-Dicke parameter. Yet, another eventual imperfection consists in a mismatch in the Rabi frequencies of the lasers forming a standing wave, i.e., $|\Omega_{r,1}-\Omega_{r,2}|=\Delta\Omega_r\neq 0$ and  $|\Omega_{b,1}-\Omega_{b,2}|=\Delta\Omega_b\neq 0$. This flaw would lead to a spurious detuned carrier, $\Delta\Omega_{r,b}/2(\sigma^+ e^{i\delta_{r,b}}+{\rm H.c.})$, which can be shown to provide approximately an effective excitation  $\sim (\Delta\Omega_{r,b})^2\sigma_z/\delta_{r,b}$. Hence, this imperfection will become significant if  $(\Delta\Omega_{r,b})^2 t/\delta_{r,b} \sim 0.1$ with $t$ the total evolution time. For the considered parameters, $\nu=2\pi\times1.5$ MHz and $t=6$ ms, it results in $\Delta\Omega_{r,b}\sim 2\pi\times 6$ kHz, which for $\Omega_{r,b}=2\pi\times 50$ kHz corresponds to a difference between the amplitudes of the two lasers forming a standing-wave of more than $10\%$. Hence, this imperfection becomes relevant here if $\Delta\Omega_{r,b}/\Omega_{r,b}\gtrsim 0.1 $. However, since better conditions have been experimentally achieved, $\Delta\Omega_{r,b}/\Omega_{r,b}\approx 0.07$ in~\cite{deLaubenfels:15}, and because the numerical simulations involving a similar trapped-ion setup performed in~\cite{Puebla:17} did not show any significant impact with $\Delta\Omega_{r,b}/\Omega_{r,b}=0.08$, we have not included this potential imperfection  in  the numerical simulations presented here.

\subsection{Numerical results}
The numerical simulations of trapped-ion dynamics in this work were performed assuming only the optical RWA, that is, starting from the following Hamiltonian
\begin{equation}
\label{eq:HTIsim}
H_{\rm TI}^I=\frac{\xi(t)}{2}\sigma_z+\sum_j \frac{\Omega_j}{2}\left[\sigma^+e^{i\eta_j \left(ae^{-i\nu t}+\adag e^{i\nu t} \right)}e^{i(\omega_I-\omega_j)t}e^{-i\phi_j}+\rm{H.c.} \right],
\end{equation}
using both bare and protected scheme with the corresponding parameters, as explained previously. Besides  magnetic field fluctuations,  for the DD scheme we include an additional noise source, namely, amplitude fluctuations of the carrier tone, i.e., $\Omega_c(1+\beta(t))$ as discussed previously.  Therefore, the results of trapped-ion simulations correspond to an ensemble average over $100$ stochastic trajectories.  In order to quantify agreement between the standard and DD scheme, we compute the state fidelity $F(t)=\left|\left<\psi_{\rm QRM}(t) \right|\left.\psi(t) \right>\right|$ with $\ket{\psi_{\rm QRM}(t)}$ the evolved quantum state under the aimed QRM,  and $\ket{\psi(t)}$ the quantum state of the trapped-ion setup obtained from Eq.~(\ref{eq:HTIsim}).

The considered trapped-ion parameters are $\nu=2\pi\times 1.5$ MHz, $\eta=0.04$, Rabi frequencies for red- and blue-sidebands $\Omega=2\pi\times50$ kHz, and a carrier with $\eta_c=0.01$ and $\Omega_c=2\pi\times 200$ kHz. In Fig.~\ref{fig:1} we show the results corresponding to a QRM for two cases with  $\omega=0$, namely, $\lambda=\omega_0$  and $\lambda=1.25 \ \omega_0$ (DSC regime) with initial states $\ket{\psi(0)}=\ket{g}\ket{0}$ and  $\ket{\psi(0)}=1/\sqrt{2}(\ket{e}-\ket{g})\ket{2}$, respectively. We recall that the initial states are referred to the spin basis of the QRM in Eq.~(\ref{eq:QRM}), and thus, they must be rotated accordingly for the protected scheme, see Eq.~(\ref{eq:HSW}) and the illustration given in Fig.~\ref{fig:schemeSW}. The simulated coupling constant results in $\lambda=\eta\Omega/2=2\pi\times 1$ kHz, and then, the evolution time goes up to $6$ ms which is $2T_2$. The results obtained by means of the DD scheme include amplitude fluctuations with a relative strength of $\zeta=5\times10^{-4}$.  The protected scheme improves the realization of the aimed QRM dynamics in the DSC regime, provided amplitude fluctuations are not too strong, as shown in Fig.~\ref{fig:1}(d). Certainly, our DD method is expected to provide similar results as the bare realization when  $P_{\rm AF}\approx P_{\rm MF}$, which corresponds to $\zeta\approx 10^{-3}$  for the considered parameters. Note that this estimation  agrees with the results of the numerical simulations plotted in Fig.~\ref{fig:1}(d). Remarkably, thanks to the magnetic-dephasing noise resilience of the dynamical decoupling scheme, for weaker amplitude fluctuations the DD method surpasses the performance of the standard scheme.

\section{Summary}
\label{sec:sum}
We propose a scheme to attain a faithful realization of the QRM in the DSC regime with a trapped ion based on continuous dynamical decoupling technique to cope with magnetic-dephasing noise. This noise constrains the coherence time, being the main source of decoherence in trapped-ion setups which involve magnetically sensitive internal states as another noise sources are expected to have a significant impact at longer evolution times. This is indeed the case of experiments performed with qubits encoded in metastable states of optical ions. This magnetic-dephasing noise, which features a finite spectral width, can be handled with dynamical decoupling techniques. However, the range of the simulated parameters  relying on these schemes is typically narrowed, and thus, tunability is traded for noise resilience.  Here, we propose a strategy to achieve a tunable QRM and at the same time robust against magnetic-dephasing noise. We demonstrate that the proposed scheme allows to explore the DSC regime of the QRM, whose main dynamical hallmark consists in the structured dynamics of collapses and revivals.

While the standard simulation of the QRM in a trapped ion is attained by means of two lasers driving detuned red- and blue-sidebands,  the proposed continuous dynamical decoupling involves the driving of a carrier interaction. This latter driving plays a fundamental role in the scheme as it defines a new dressed basis to encode the qubit and handles magnetic-dephasing noise. Nevertheless, under typical trapped-ion parameters, unwanted terms may deteriorate the correct simulation of the QRM as a consequence of an additional RWA. This obstacle can be overcome by properly setting red- and blue-sideband lasers in a standing wave configuration. Despite the proposed scheme is resilient against magnetic-dephasing noise, another potential imperfections may challenge a better performance of this scheme. Indeed, fluctuations in the laser amplitude  lead to dephasing noise in the dressed basis in which the dynamical decoupling method operates, and thus, the strength of this noise sets an additional constraint for the correct functioning of the proposed method.

\section*{Funding}
This work was supported by the ERC Synergy grant BioQ, the EU STREP
project EQUAM and the CRC TRR21. The authors acknowledge support by the state of Baden-W\"urttemberg through bwHPC and the German Research Foundation (DFG) through grant no INST 40/467-1 FUGG. J. C. acknowledges Universit\"at Ulm for a Forschungsbonus.

\bibliographystyle{tfp}

\end{document}